\documentclass[preprint,12pt]{elsarticle}

\usepackage{amssymb}
\usepackage{amsmath}
\usepackage{tabularx}
\usepackage{stackengine}
\usepackage{xcolor}

\journal{CAF}

\begin{document}

\begin{frontmatter}



\title{A flux bounce-back scheme for the filtered Spectral Element Lattice Boltzmann Method}


\author[a]{Chunheng Zhao}
\author[b]{Saumil S. Patel}
\author[a]{Hai Lu Lin}
\author[c]{Misun Min}
\author[a]{Taehun Lee}

\address[a]{Department of Mechanical Engineering, City College of New York, New York, NY 10031, USA}
\address[b]{Computational Science Division, Argonne National Laboratory, Lemont, IL 60439, USA}
\address[c]{Mathematics and Computer Science Division, Argonne National Laboratory, Lemont, IL 60439, USA}

\begin{abstract}
We develop a spectral element lattice Boltzmann method (SELBM) with the flux bounce-back (FBB) scheme, to enable accurate simulations of single-phase fluid dynamics in unstructured mesh. We adopt an Eulerian description of the streaming process in place of the perfect shift in the regular LBM. The spectral element method is used to spatially discretize the convective term, while the strong stability-preserving Runge-Kutta (SSPRK) method is used for time integration. To increase stability, we investigate the use of an explicit filter, particularly in the context of the sensitive double shear layer problem. The results indicate that by using the high-order polynomial, we can effectively eliminate the small vortices around the neck region. We introduce the flux bounce-back scheme to enable the current scheme to handle complex boundaries. The proposed scheme and flux boundary method are validated through benchmark simulations, including the unsteady Couette flow and the planar Poiseuille flow. Further validation is provided through the Taylor-Green vortex problem, demonstrating the accuracy and convergence of the scheme for isotropic turbulence. Finally, we consider a fully developed turbulent flow within a cylindrical pipe and correctly predict the turbulent boundary layer profile.

\end{abstract}


\begin{highlights}
\item Spectral Element method
\item Lattice Boltzmann method
\item Flux bounce-back scheme
\item Turbulent pipe flow
\end{highlights}


\end{frontmatter}


\section{Introduction}

The lattice Boltzmann method (LBM) is widely used as a numerical solver for fluid dynamics in single- and multi-phase flows~\cite{kruger2017lattice}. Instead of directly solving the macroscopic governing equations (i.e. Navier-Stokes equations), LBM solves the evolution of the particle distribution function and recovers density and momentum through moments of the particle distribution functions~\cite{mohamad2011lattice}. It is computationally efficient due to a local collision function and a perfect shift streaming function~\cite{noble1996comparison,zhao2025difference}. These features enable massive parallelization, making LBM particularly well suited for modern high-performance computing architectures~\cite{zhao2025imexlbm}. The perfect-shift approach to the streaming step in the LBM ensures that the distribution function propagates along the characteristic direction with a constant speed, effectively eliminating numerical errors in the advection step. As a result, the primary source of error in LBM arises from the collision step, which maintains second-order spatial accuracy. Through Chapman-Enskog analysis, the Navier-Stokes equations can be recovered and directly derived from LBM~\cite{he1997lattice,junk2005asymptotic,dubois2008equivalent}. Although the LBM provides a relatively straightforward alternative to directly solving the Navier-Stokes equations, several critical aspects need careful consideration. Typically, LBM is implemented on a structured mesh~\cite{lee2003eulerian}, or lattice grid, where the spatial interval $\Delta x $ and the time step $\Delta t$ are set to one lattice unit, and the Courant-Friedrichs-Lewy (CFL) number is fixed at $1$, implying a uniform grid-spacing. In certain scenarios, unstructured meshes are indispensable. For example, boundary layer meshes are required to accurately capture the turbulent boundary layer profile~\cite{garimella2000boundary}. Using a regular LBM scheme in such cases would require excessive grid refinement, leading to a significant waste of computational resources. Other examples arise in simulations involving complex geometries, such as fluid flow in porous media or pebble bed reactors~\cite{perkins1963review,jiang2019review}.

The LBM on unstructured meshes can be broadly categorized into finite volume (FV) LBM~\cite{xi1999finite,li2016finite,mishra2007solving,patil2009finite} and finite element (FE) LBM~\cite{lee2003eulerian,min2011spectral,lee2001characteristic,saadat2020arbitrary}. In unstructured meshes, wall normals generally vary across adjacent boundary surfaces, making it difficult to clearly define the incoming and outgoing directions. As a result, identifying the correct bounce-back directions for boundary conditions in unstructured meshes becomes particularly challenging~\cite{reis2012lattice,reis2020lattice,zhang2012general}. Moreover, the perfect shift in the streaming used on structured meshes cannot be directly applied to unstructured meshes, which may introduce additional stability and accuracy issues. In finite volume LBM, modified bounce-back schemes~\cite{patil2009finite} and distance-based bounce-back approaches~\cite{li2018gas} have been proposed to address complex boundaries. However, most finite volume formulations directly discretize the discrete Boltzmann equation~\cite{deville2002high}, which may induce stability concerns at high Reynolds numbers. Finite element LBM provides a more natural framework for constructing high-order streaming operators, such as those based on spectral elements. For the discontinuous Galerkin (DG) LBM, flux-based boundary conditions are applied at the boundary element interfaces~\cite{min2011spectral}. For the continuous Galerkin LBM~\cite{lee2003eulerian}, flux boundary conditions combining with a bounce-back scheme can be imposed. However, further investigation is needed to provide a physical interpretation for the Lax-Wendroff-type boundaries~\cite{lee2003eulerian,lee2001characteristic}, which involve high-order streamline diffusion. The DG-LBM has demonstrated the ability to produce stable solutions using the flux bounce-back (FBB) scheme. However, this approach comes with a trade-off: the DG method inherently introduces additional dissipation at the element interface to enhance stability at high Reynolds numbers. Building on previous research, we have developed a spectral element lattice Boltzmann method based on a continuous Galerkin formulation. Unlike the Lax-Wendroff streaming method, our approach employs a Strong Stability-Preserving (SSP) Runge-Kutta scheme to minimize numerical dissipation. Additionally, a non-dissipative filter is applied to the distribution functions, effectively preserving both density and momentum conservation as well as maintaining numerical stability. 

The following sections present the proposed scheme and its validation through benchmark simulations. Section 2 describes the methodology and formulation of the scheme. Section 3 provides validation results through a series of benchmark tests, including the stability test, double shear layer, boundary condition verification, unsteady Couette flow, and planar Poiseuille flow. Additional validations are performed for turbulent flows using the Taylor Green vortex and the Poiseuille pipe flow.

\section{Methodology}
\label{sec:Methodology}
\subsection{Lattice Boltzmann method}

We start with the discrete Boltzmann equation of the density distribution function $f_\alpha$ for the $\alpha $-direction, with an external forcing term $F_\alpha$:
\begin{equation}\label{dbe_m}
    \left(\frac{\partial}{\partial t}+\mathbf{e}_\alpha\cdot\nabla\right)f_\alpha=-\frac{1}{\lambda}(f_\alpha-f_\alpha^{eq})+F_\alpha,
\end{equation}
where $f_\alpha^{eq}$ denotes the equilibrium distribution function given as:
\begin{equation}\label{geq_s}
    f_\alpha^{eq}=t_\alpha\rho\left(1+ \frac{\boldsymbol{e}_\alpha \cdot \mathbf{u}}{c_s^2}+
\frac{(\boldsymbol{e}_\alpha\cdot\boldsymbol{u})^2}{2c_s^4}-
\frac{|\boldsymbol{u}|^2}{2c_s^2} \right).
\end{equation}
In Equations~(\ref{dbe_m}) and~(\ref{geq_s}), the weights $t_\alpha$ are associated with the equilibrium distribution function. The vector $\boldsymbol{e_\alpha}$ represents the discrete velocity vector for $\alpha$-direction. The parameter $\lambda$ represents the relaxation time, which is connected to the kinematic viscosity $\nu$ through the equation $\nu=\lambda c_s^2$, where $c_s$ denotes the speed of sound. For this study, we chose the $D2Q9$ and $D3Q19$ models, and the parameters mentioned above are given in~\cite{zhao2023interaction,lee2005stable}. The external forcing term can be expressed as:
\begin{equation}
    F_\alpha=t_\alpha\left(\frac{\mathbf{e}_\alpha-\boldsymbol{u}}{c_s^2}+\frac{(\boldsymbol{e}_\alpha\cdot\boldsymbol{u})\boldsymbol{e}_\alpha}{c_s^4}\right)\cdot\boldsymbol{F}_\alpha,
\end{equation}
which can be expressed as a combination of a leading-order term and a higher-order terms~\cite{patel2016new}. The leading-order forcing term $F_\alpha^*$ can be expressed as:
\begin{equation}\label{body}
    F^*_\alpha=t_\alpha\left(\frac{\mathbf{e}_\alpha}{c_s^2}\right)\cdot \boldsymbol{F}_\alpha,
\end{equation}
and the higher-order forcing term $F_\alpha^{**}$ is then:
\begin{equation}\label{f_c}
    F^{**}_\alpha=t_\alpha\left(\frac{(\boldsymbol{e}_\alpha\cdot\boldsymbol{u})\boldsymbol{e}_\alpha}{c_s^4}-\frac{\boldsymbol{u}}{c_s^2}\right)\cdot\boldsymbol{F}_\alpha.  
\end{equation}
Here, the body force is represented by $\boldsymbol{F}_\alpha$. We then integrate Eq.~(\ref{dbe_m}) along the characteristic direction and apply the trapezoidal rule. In addition, we introduce the modified equilibrium distribution function as follows~\cite{lee2006}.
\begin{equation}
    \bar{f}_\alpha^{eq}=f_\alpha^{eq}-0.5\Delta t F_\alpha^{**},
\end{equation}
and the modified distribution:
\begin{equation}
    \bar{f}_\alpha=f_\alpha+\frac{f_\alpha-f_\alpha^{eq}}{2\tau}-0.5\Delta t F_\alpha^{**},
\end{equation}
where $\tau=\lambda/\Delta t$ denotes the dimensionless relaxation time.
The discrete Boltzmann equation can be then solved by a local collision function and a non-local streaming function with the force splitting method~\cite{patel2016new}. The collision is conducted as follows:
\begin{equation}
    f_\alpha^*=\bar{f}_\alpha-\frac{1}{\tau+0.5}(\bar{f}_\alpha-\bar{f}_\alpha^{eq})+\Delta t F_\alpha^{**},
\end{equation}
which is followed by the substitution:
\begin{equation}\label{perfect1}
    \bar{f}_\alpha(\boldsymbol{x},t-\Delta t)= f^*_\alpha(\boldsymbol{x},t-\Delta t),
\end{equation}
and the streaming is given as:
\begin{equation}\label{advection0}
    \left(\frac{\partial}{\partial t }+\boldsymbol{e}_\alpha\cdot\nabla\right)\bar{f}_\alpha =F_\alpha^*.
\end{equation}
Finally, the density and momentum can be recovered by the distribution function by: 
\begin{equation}\label{pressure}
    \rho=\sum_\alpha \bar{f}_\alpha,
\end{equation}
\begin{equation}\label{momentum}
    \rho\boldsymbol{u}=\sum_\alpha \boldsymbol{e}_\alpha \bar{f}_\alpha .
\end{equation}

\subsection{Spectral Element Continuous Galerkin method}

By the force splitting method~\cite{patel2016new}, we can incorporate the higher-order forcing term $F_\alpha^{**}$ into the collision operator to preserve the conserved moments, i.e., $\sum_\alpha F_\alpha^{**}= \sum_\alpha \boldsymbol{e}_\alpha F_\alpha^{**} = 0$, while the leading-order forcing term $F_\alpha^*$ is applied to the streaming step. We express the streaming process as a pure advection equation of the particle distribution function based on an unstructured mesh shown in Eq.~(\ref{advection0}), instead of the perfect shift.
The weak form of Eq.~(\ref{advection0}) in a given element domain $\Omega_e$, can be obtained by multiplying the advection equation Eq.~(\ref{advection0}) by a test function $\phi$:
\begin{equation}\label{advection}
    \left(\frac{\partial\bar{f}_\alpha}{\partial t }+\boldsymbol{e}_\alpha\cdot\nabla \bar{f}_\alpha -F_\alpha^*,\phi\right)_{\Omega_e}=0.
\end{equation}
We can express the above equation in the following matrix form:
\begin{equation}\label{full}
    \boldsymbol{M}\frac{d\boldsymbol{\bar f}_\alpha}{d t }=-\boldsymbol{C}_\alpha\boldsymbol{\bar f}_\alpha+\boldsymbol{M}\boldsymbol{F}_\alpha^*,
\end{equation}
where $\boldsymbol{M}$ and $\boldsymbol{C_\alpha}$ are the mass and convection matrices respectively, whose entries are given as:
\begin{equation}\label{mass}
    {M}_{ij}=\int_{\Omega_e}\phi_i\phi_j\ d\Omega,
\end{equation}
\begin{equation}\label{advection2}
    {C}_{\alpha_{ij}}=\sum_{k=1}^d\left(e_{\alpha,x_k}\int_{\Omega_e}\phi_i\partial_{x_k}\phi_j\ d\Omega\right).
\end{equation}
The particle distribution function and the forcing terms defined on the nodes become vectors or one-dimensional arrays for sequential storage, shown as $\boldsymbol{\bar f}_\alpha$ and $\boldsymbol{F}_\alpha^*$. We employ a tensor product basis of the 1D Legendre–Lagrange interpolation polynomials given as:
\begin{equation}
    l_i(\xi)=\frac{-1}{N(N+1)}\frac{(1-\xi^2)L'_N(\xi)}{(\xi-\xi_i)L_N(\xi_i)}
\end{equation}
where $\xi_i$ are the Gauss-Lobatto-Legendre (GLL) quadrature nodes. Based on GLL quadrature, we can obtain the formulation for lumped diagonal mass matrix in $3D$ as:
\begin{equation}
    \boldsymbol{M}[\hat{i},\hat{i}]=\phi_{\hat{i}}^2J(\boldsymbol{r})\rho_i\rho_j\rho_k, \ \hat{i}=1+i+(N+1)j+(N+1)^2k
\end{equation}
where 
$$\rho_i=\frac{2}{N(N+1)}\frac{1}{[L_N(\xi_i)]^2}$$
represents the specific weights, and $J(\boldsymbol{r})$ denotes the Jacobian determinant according to the reference domain where:
\begin{equation}
J(\boldsymbol{r})=det\frac{\partial x_i}{\partial r_i}
\end{equation}
The convection matrix $\boldsymbol{C}_\alpha$ can be expressed as:
\begin{equation}
\boldsymbol{C}_\alpha=\underline{\phi}^T\boldsymbol{e}_\alpha\boldsymbol{D}\underline{\phi},
\end{equation}
where 
\begin{equation}
\boldsymbol{D}=\left\{
\begin{aligned}
&(\boldsymbol{D}_x,\boldsymbol{D}_y)^T, & d=2; \\
&(\boldsymbol{D}_x,\boldsymbol{D}_y,\boldsymbol{D}_z)^T, &d=3,
\end{aligned}
\right.
\end{equation}
is the derivative matrix composed of $d$ matrices and $\boldsymbol{D}_x$ is composed of the one-dimensional derivative matrix $\boldsymbol{\hat{D}}_x$ related to GLL points. When we consider a two-dimensional case, 

\begin{equation}
\begin{aligned}
&\boldsymbol{D}_x=\boldsymbol{I}\otimes\boldsymbol{\hat{D}}_x, \\
&\boldsymbol{D}_y=\boldsymbol{\hat{D}}_y\otimes\boldsymbol{I},
\end{aligned}
\end{equation}
where $\boldsymbol{I}$ is the identity matrix, and the operator $\otimes$ denotes the tensor product~\cite{fischer2022nekrs}. For a three-dimensional simulation, 

\begin{equation}
\begin{aligned}
&\boldsymbol{D}_x=\boldsymbol{I}\otimes\boldsymbol{I}\otimes\boldsymbol{\hat{D}}_x, \\
&\boldsymbol{D}_y=\boldsymbol{I}\otimes\boldsymbol{\hat{D}}_y\otimes\boldsymbol{I},\\
&\boldsymbol{D}_z=\boldsymbol{\hat{D}}_z\otimes\boldsymbol{I}\otimes\boldsymbol{I}.
\end{aligned}
\end{equation}
After we transfer the domain to the reference domain, from $[x,y,z]$ to $[r,s,t]$, the chain rule needs to be considered and the change of variables should be done from $\Omega$ to $\hat{\Omega}$:
\begin{equation}
    \boldsymbol{C}_\alpha=\underline{\phi}^T\boldsymbol{e}_\alpha\boldsymbol{G}\boldsymbol{D}^*\underline{\phi},
\end{equation}
where now
\begin{equation}
\boldsymbol{D}^*=\left\{
\begin{aligned}
&(\boldsymbol{D}_r,\boldsymbol{D}_s)^T, & d=2; \\
&(\boldsymbol{D}_r,\boldsymbol{D}_s,\boldsymbol{D}_t)^T, &d=3,
\end{aligned}
\right.
\end{equation}
The geometry information is then stored in the matrix $\boldsymbol{G}$, which is composed of $d^2$ diagonal matrices:
\begin{equation}
\boldsymbol{G}=
\begin{aligned}
\begin{pmatrix}
\boldsymbol{G}_{11} & \boldsymbol{G}_{12} \\
\boldsymbol{G}_{21} & \boldsymbol{G}_{22}
\end{pmatrix},  &d=2,\\
\end{aligned}
\end{equation}

\begin{equation}
\boldsymbol{G}=
\begin{aligned}
\begin{pmatrix}
\boldsymbol{G}_{11} & \boldsymbol{G}_{12}  &\boldsymbol{G}_{13}\\
\boldsymbol{G}_{21} & \boldsymbol{G}_{22}  &\boldsymbol{G}_{23}\\
\boldsymbol{G}_{31} & \boldsymbol{G}_{32}  &\boldsymbol{G}_{33}
\end{pmatrix},  &d=3.\\
\end{aligned}
\end{equation}
The diagonal matrix can then be expressed as
\begin{equation}
    \boldsymbol{G}_{ij}[\hat{l},\hat{l}]=\rho_l\rho_m\rho_n\sum_{k=1}^d\frac{\partial r_i}{\partial x_k}J(\boldsymbol{r}),
\end{equation}
where $\hat{l}=1+l+(N+1)m+(N+1)^2n$.

\subsubsection{Flux Bounce-Back Boundary Condition}

From the scenario of the discontinuous Galerkin method, it is intuitive to numerically compute the flux at the interface as well as the boundary condition. Here, in our continuous Galerkin LBM, we weakly impose the boundary condition only at the boundary nodes, for which the bounce back scheme is implicitly applied~\cite{min2011spectral}. As shown in previous work~\cite{min2011spectral}, we further introduce a surface integration of the flux term from the outer boundary rather than each element interface, and the entire equation can be expressed as:
\begin{equation}\label{entire}
    \boldsymbol{M}\frac{\partial\boldsymbol{\bar f}_\alpha}{\partial t}+\boldsymbol{C}_\alpha\boldsymbol{\bar f}_\alpha=\boldsymbol{M}\boldsymbol{F}^*_\alpha+\boldsymbol{R}(\boldsymbol{n}\cdot\boldsymbol{j}_\alpha),
\end{equation}
where $\boldsymbol{R}=\int_{\partial\Omega_e} \phi_i\phi_j\   d\bar{\Omega}$ is the surface integration, $\boldsymbol{n}$ is the normal vector, and $\boldsymbol{j}_\alpha$ is the flux at the boundary. In this work, we apply the Lax-Friedrichs flux:
$$ \boldsymbol{j}_\alpha=\left\{
\begin{aligned}
&\boldsymbol{e}_\alpha[\boldsymbol{\bar f}_\alpha]_{bc},&     &\boldsymbol{n}\cdot\boldsymbol{e}_\alpha <0, \\
&\boldsymbol{0} & & \boldsymbol{n}\cdot\boldsymbol{e}_\alpha \ge0.
\end{aligned}
\right.
$$
When we apply the bounce-back at the boundary, the flux term is expressed as:
\begin{equation}
    [\boldsymbol{\bar f}_\alpha]_{bc}=\boldsymbol{\bar f}_\alpha-\boldsymbol{\bar f}_\beta-2t_\alpha\rho_0(\boldsymbol{e}_\alpha\cdot \boldsymbol{u}_b)/c_s^2,
\end{equation}
for which $\beta$ is the opposite direction of $\alpha$. When no slip is applied to the simulation, we simply apply flux bounce-back at the boundary. When we have a moving boundary with velocity $\boldsymbol{u}_b$, we further add the last term to maintain a constant velocity. 

It is noted that the integration of the flux term is applied only to the distribution functions for incoming directions, while the other distribution functions will not be affected by this flux term. Essentially, when we have a flat surface boundary, the flux bounce-back will yield the same results as the normal bounce-back approach.   

\subsubsection{Time marching method}
We consider the Runge-Kutta to explicitly solve the distribution function. To solve the distribution function $\boldsymbol{\bar f}_\alpha $ in Eq.~(\ref{entire}), we propose the 3rd order strong stability-preserving Runge-Kutta method for:
\begin{equation}
    \frac{d\boldsymbol{\bar f}_\alpha}{d t}=\boldsymbol{L}\boldsymbol{\bar f}_\alpha,
\end{equation}
where $\boldsymbol{L}=\boldsymbol{I}+\boldsymbol{M}^{-1}\left(\boldsymbol{R}-\boldsymbol{C}_\alpha\right)$ is the spatial operator. The each turn of the SSP RK3 is then:
\begin{equation}\label{RK1}
    \boldsymbol{\bar f}^1_\alpha=\boldsymbol{\bar f}_\alpha(t-\Delta t)+\frac{\Delta t}{3}\boldsymbol{L}\boldsymbol{\bar f}_\alpha(t-\Delta t),
\end{equation}
\begin{equation}\label{RK2}
    \boldsymbol{\bar f}_\alpha^2=\boldsymbol{\bar f}_\alpha^1-\frac{3\Delta t}{4}\boldsymbol{L}( \boldsymbol{\bar f}_\alpha(t-\Delta t)- \boldsymbol{\bar f}_\alpha^1),
\end{equation}
\begin{equation}\label{RK3}
    \boldsymbol{\bar f}_\alpha^3=\boldsymbol{\bar f}_\alpha^2+\frac{2\Delta t}{3}\boldsymbol{L}(\boldsymbol{\bar f}_\alpha(t-\Delta t)- \boldsymbol{\bar f}_\alpha^1+\boldsymbol{\bar f}_\alpha^2),
\end{equation}
\begin{equation}\label{RK4  }
    \boldsymbol{\bar f}_\alpha(t)=\boldsymbol{\bar f}_\alpha^3-\frac{\Delta t}{4}\boldsymbol{L}(\boldsymbol{\bar f}_\alpha(t-\Delta t)- \boldsymbol{\bar f}_\alpha^1+\boldsymbol{\bar f}_\alpha^2-\boldsymbol{\bar f}_\alpha^3).
\end{equation}
As we get the the current time distribution function $\boldsymbol{\bar f}_\alpha(t)$, we can further evaluate the macroscopic density and momentum from Eq.~(\ref{pressure}) and (\ref{momentum}).
\subsubsection{Filter $\&$ Stabilization}
One key difference between the current scheme and the conventional LBM lies in the streaming process. The conventional LBM employs perfect shift streaming without any numerical error. As a result, the stability of the conventional LBM is typically enhanced through the collision process~\cite{ricot2009lattice}. Methods such as multi-relaxation time (MRT) collision~\cite{luo2011numerics}, two-relaxation time (TRT) collision~\cite{ginzburg2008two}, increasing bulk viscosity~\cite{dellar2001bulk}, and regularizing the collision~\cite{latt2006lattice} operator are commonly used to stabilize the LBM solver for high Reynolds number simulations. Filtering macroscopic values, such as density or velocity, can introduce conservation issues and is therefore generally avoided~\cite{ricot2009lattice}. In the current scheme, the perfect shift streaming is changed to the spectral element and RK scheme. Although the continuous Galerkin method ensures a continuous distribution function across the element interface, the gradient of the distribution function term, which appears in the convective term $\boldsymbol{e_\alpha}\cdot\nabla f_\alpha$, is not exactly continuous, which may induce stability issues. 

In Nek5000~\cite{Nek5000_repo}, when simulating a fluid system with a high Reynolds number, an explicit filter is used to suppress the spurious modes of pressure and momentum at the end of each time step. This filter interpolates the nodal solution using a pseudo-projection~\cite{deville2002high, fischer2001filter}. After each solving process, the velocity and pressure terms in space are replaced with filtered solutions by applying a filter transfer function $\sigma$. For example, in an $N^\text{th}$-order spectral element system, the filtered pressure term can be expressed as:
\begin{equation}
    p=\sum_{k=0}^N\sigma_k p_k\phi_k(\xi),
\end{equation}
where $p_k$ is the physical space solution, and $\phi_k(\xi)$ is the test function. The filter transfer function can be expressed as:
\begin{equation}
\sigma_k=\left\{
\begin{aligned}
&1-\gamma\left(\frac{k-k_c}{N-k_c}\right)^2, & k>k_c, \\
&1, & k\leq k_c.\\
\end{aligned}
\right.
\end{equation}
In the above equation, $k_c$ is defined as the cutoff mode, and $\gamma$ denotes the amplitude. For example, when we choose $k_c = 2$ and $\gamma = 0.1$, the filter is applied starting from the $2^\text{nd}$ order solution, and the interpolation magnitude for higher-order solutions is $0.1$. Further details on the construction of the filter and interpolation can be found in~\cite{deville2002high, fischer2001filter}. However, it is important to note that, although this is a non-dissipative method, it violates the divergence-free condition for incompressible flow when we use the explicit filter on the velocity and pressure in Nek5000.

For our approach, we do not apply the explicit filter to the density and momentum, as this would violate the conservations $\sum f_\alpha^{eq} = \sum f_\alpha$ and $\sum f_\alpha^{eq} \boldsymbol{e}_\alpha = \sum f_\alpha \boldsymbol{e}_\alpha$. Instead, the explicit filter is directly applied to the distribution function $f_\alpha$ before the update step. In this case, conservation relations are preserved.





 We summarize the full process of the scheme as:

$\boldsymbol{1.}$ Initialize of the distribution on each node and element;

$\boldsymbol{2.}$ Compute the loop from the collision by BGK, with higher order forcing term;

$\boldsymbol{3.}$ Solve streaming as a hyperbolic partial differential equation with external lower order forcing term and boundary flux by SSP RK3;

$\boldsymbol{4.}$ Apply the filter for each direction distribution function;

$\boldsymbol{5.}$ Update the macroscopic parameters which are essential for the next collision step.

\section{Simulation Results}
The validations are performed for the investigation of the stability and boundary conditions. Our results are generated from a modified version of the Nek5000 code, which we henceforth refer to as NekLBM. NekLBM leverages many of the numerical operators and parallelization inherent in Nek5000 with the distinguishing feature of solving the discrete Boltzmann equations rather than the Navier-Stokes equations. 

\subsection{Double shear layer}
\begin{figure}
	\centering
  \includegraphics[width=\linewidth]{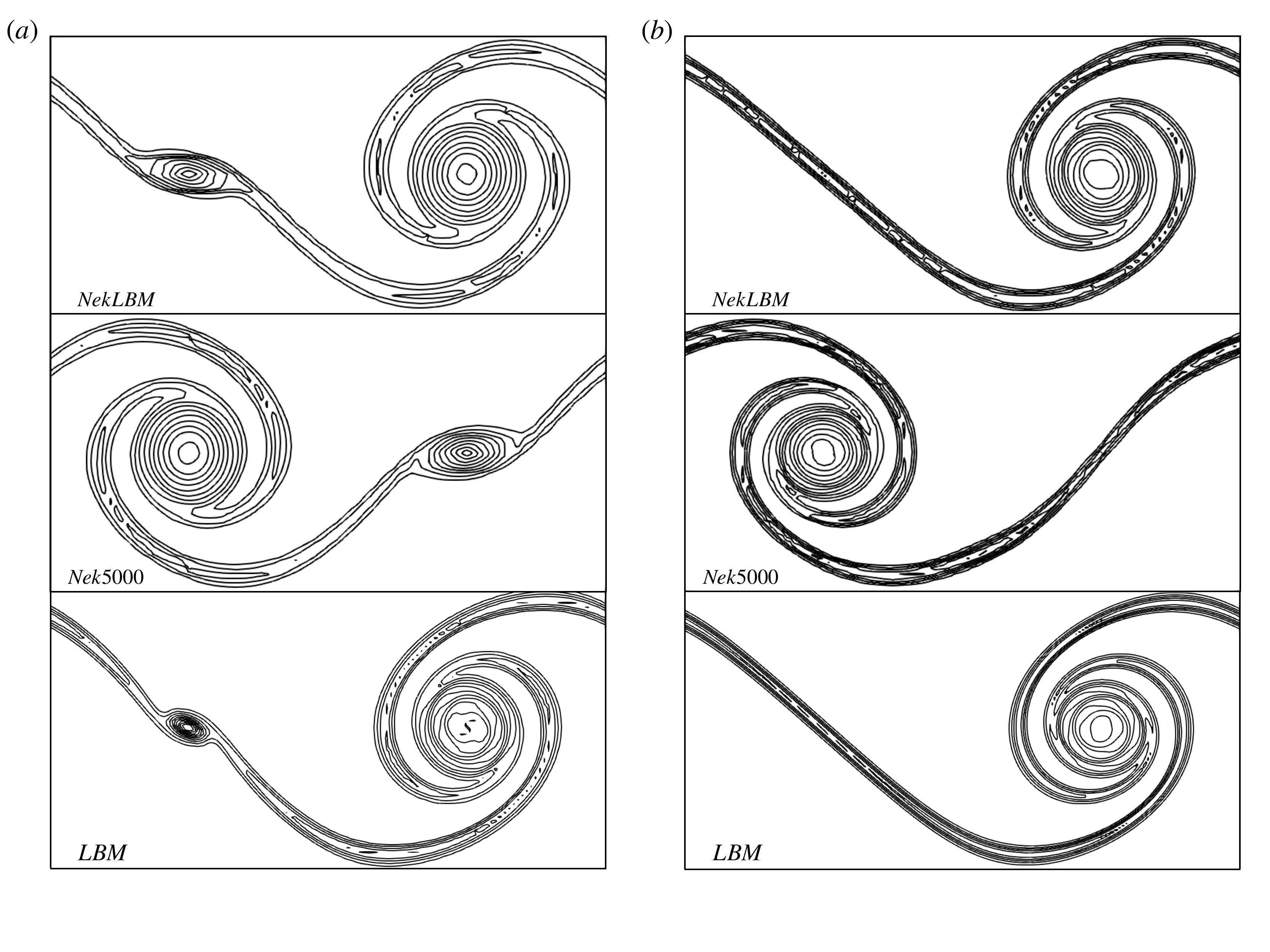}
    \caption{\label{compare} Comparison of the double shear layer simulation conducted by NekLBM, Nek5000, and regular LBM. (a) Vortex contours $\Delta \omega=5$ for $N_e=32$, $N=4$ by NekLBM(top), Nek5000 (middle), and regular LBM with $128^2$ number of nodes. (b) Vortex contours $\Delta \omega=5$ for $N_e=16$, $N=8$ by NekLBM(top), Nek5000 (middle), and regular LBM with $256^2$ number of nodes.
    }
\end{figure}

The first validation case considered is the periodic double shear layer simulation, which is studied to evaluate the solver's stability. In previous LBM studies, Dellar increased the bulk viscosity to effectively resolve small vortices induced by instability~\cite{dellar2001bulk}, while Ricot et al. investigated various types of filters applied to the distribution function to address the problem of small vortices~\cite{ricot2009lattice}. In this work, it is necessary to revisit this problem using the current model.

We initialize a velocity profile:
\begin{equation}
u_x=\left\{
\begin{aligned}
&u_0 \tanh[\kappa(y-0.25)], & y\le0.5, \\
&u_0 \tanh[\kappa(0.75-y)], & y>0.5,\\
\end{aligned}
\right.
\end{equation}
$$u_y=\delta \sin(2\pi(x+0.25))$$
in a square domain. The initial velocity magnitude $u_0$ is set to satisfy $Re=\rho u_0 L/\mu =10,000$, and the parameters related to the velocity profile are given as $\delta=0.05$, $\kappa =80$. The simulation is conducted over the time interval $T/t_0=[0,10]$, where the time scale is $t_0=L/u_0$ and we compare the results from our scheme with those from Nek5000 using the same setup at $T/t_0=10$, as shown in Figure~\ref{compare}. In Figure~\ref{compare}(a), we fix the number of elements $N_e=32$ on each direction, and set the polynomial order $N=4$ for both methods. The small vortices appear in both simulation results because of the inaccurate evaluation of the convective term. When we increase the polynomial order to $N=8$ while keeping the total number of nodes in each direction $N_{tot}=128$, our scheme produces better results, which are free from the small vortices, and consistent with Nek5000. As the polynomial order increases, the evaluation of the convective term $\boldsymbol{e_\alpha}\cdot \nabla f_\alpha$ is more accurate in space. Although the total number of nodes is the same, a higher order simulation will generally provide more accurate results. It is worth noting that for both schemes, an explicit filter with amplitude $\gamma=0.1$, and cutoff mode $k_c=2$ is applied. Without the explicit filter, both schemes fail to produce meaningful results.
\subsection{Unsteady Couette flow}
\begin{figure}
	\centering
  \includegraphics[width=\linewidth]{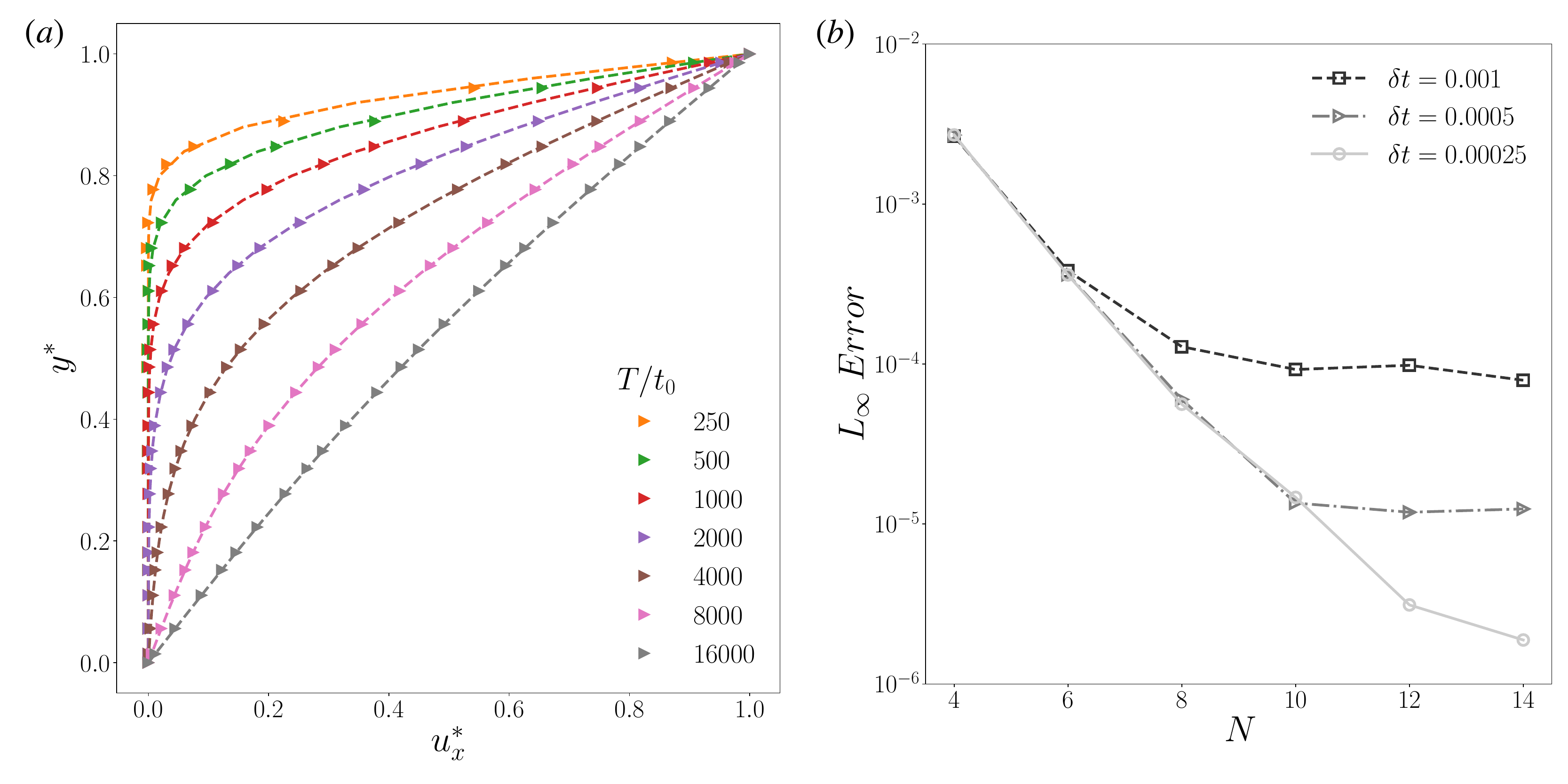}
    \caption{\label{converge1} (a) Comparison between simulation results (triangular markers) and analytical solutions (dashed line) of the unsteady Couette flow during $T/t_0=[0.5-64]$ for $Re=10$, $Ma=0.1$. (b) Spatial convergence for the unsteady Couette flow with different $\delta t$ for $Re=2000$, $Ma=0.05$.
    }
\end{figure}
A convergence test for 2D Couette flow simulation is investigated to validate the boundary condition for the current method. We initialize a moving wall at the top of the square domain, and keep a stationary no-slip boundary on the bottom. The left and right boundary conditions are set to periodic. We first test the case with $Re=u_{top}L/\nu =10$ and $Ma=u_{top}/c_s=0.1$. The viscous stress will induce the flow over the domain. For the parameters, the number of elements in each direction is set to $N_e=6$ and the polynomial order is set to $N=4$. The simulation results are compared with the analytic solutions shown in Figure~\ref{converge1}(a) and consistent results are obtained.

We further increase the Reynolds number to $Re=2000$, and set $Ma=0.05$. We evaluate the $L_\infty$ errors at $T/t_0=40$, where $t_0=L/u_{top}$, and conduct the convergence test as we increase $N$ from $N=4$ to $N=14$. As shown in Figure~\ref{converge1}(b), the errors
show exponential convergence as $N$ increases until it is saturated by temporal error.

\subsection{Planar Poiseuille flow}
\begin{figure}
	\centering
  \includegraphics[width=\linewidth]{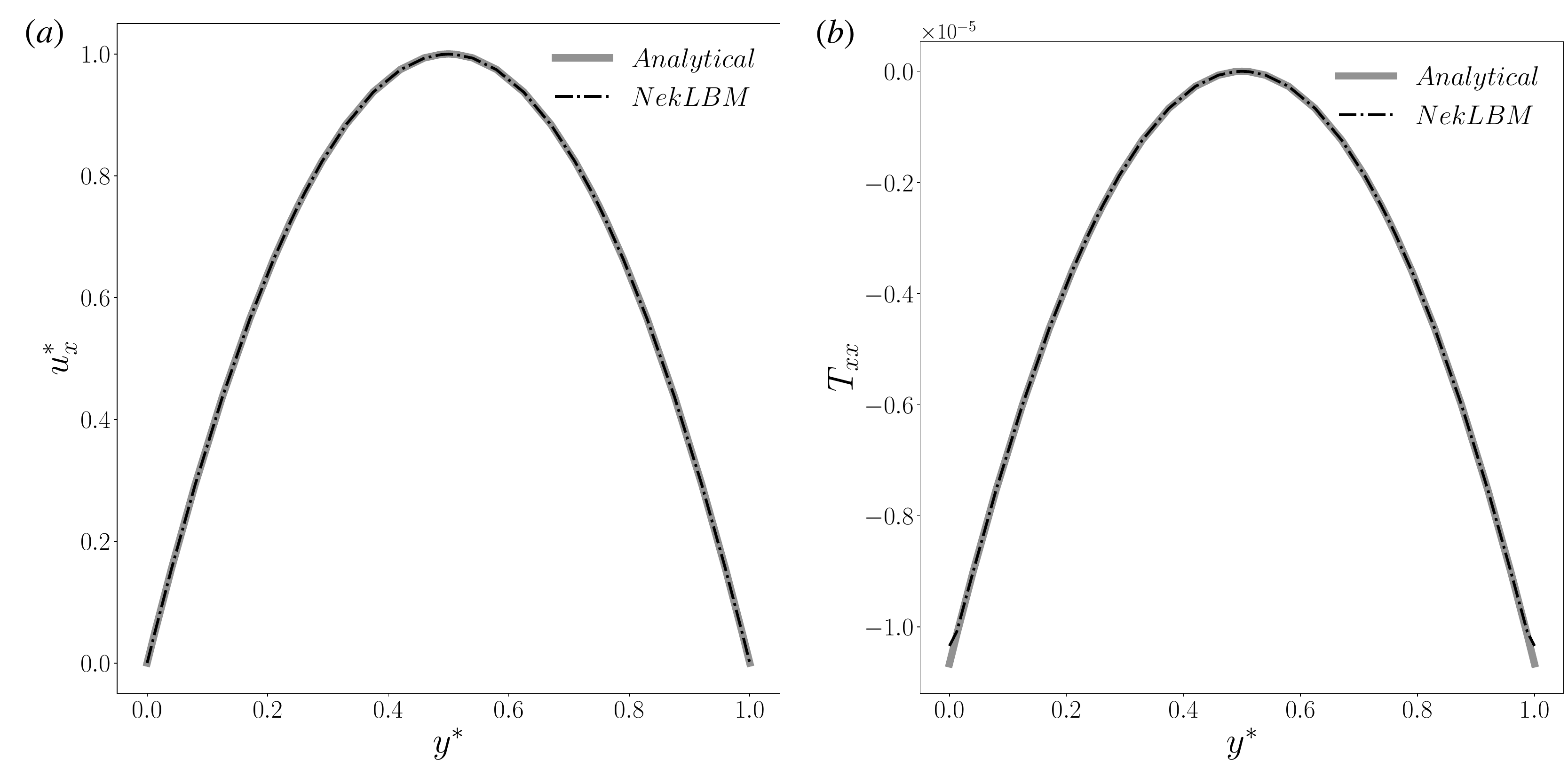}
    \caption{\label{converge} Comparison between simulation results (black dashed line) and analytical solutions (gray) of (a)velocity and (b) stress for
    the 2D planar Poiseuille flow at $T/t_0=64$ with $Ma=0.1/\sqrt{3}$, $Re=100$.
    }
\end{figure}
\begin{figure}
	\centering
  \includegraphics[width=\linewidth]{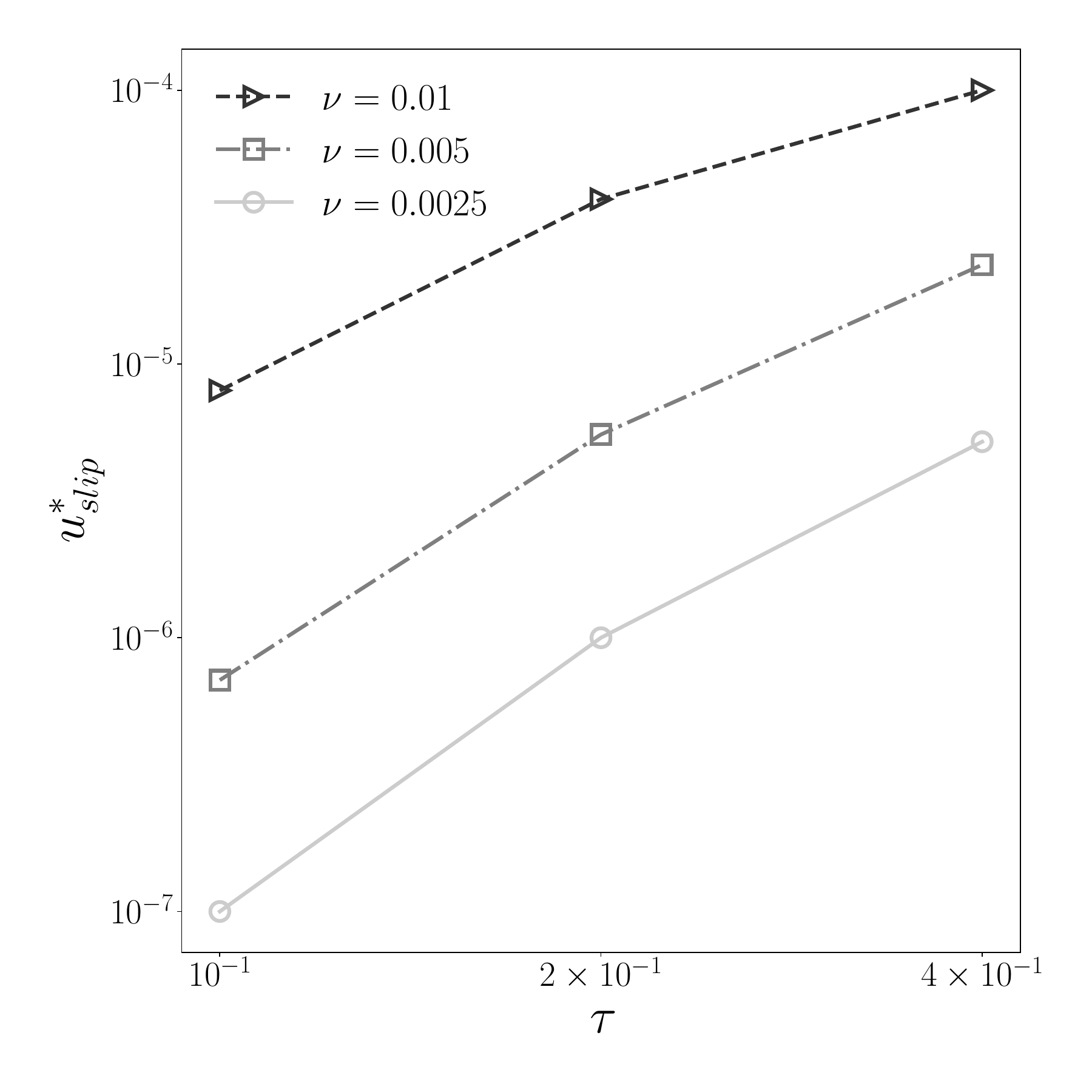}
    \caption{\label{slip} Convergence test of the dimenionless slip velocity $u_{slip}^*=u_x(y=0)/u_{max}$ with relaxation time $\tau=[0.1,0.4]$, and kinematic viscosity $\nu=[0.0025,0.01]$.
    }
\end{figure}
We further test the current model by simulating a 2D planar Poiseuille flow to study the "numerical slip" and the effect of implementing the flux boundary condition. The nonphysical "numerical slip" at the boundary of solids cannot be fully removed by node-based bounce-back method~\cite{he1997analytic,ginzbourg1994boundary}. We can improve the method by an improved collision operator i.e. MRT/TRT or coupling linked-based bounce-back with momentum exchange boundary to recover exact solutions~\cite{cercignani1989kinetic,ladd1994numerical}. The current boundary condition considers the momentum exchange through adding the flux term. Moreover, this method can highly improve the stability for unstructured mesh boundaries.

The simulation is conducted in a square domain with $N_e=4$ and $N=8$. The initial velocity is set to $\boldsymbol{u}=\boldsymbol{0}$. To drive the flow in the confined planar, we initialize a pressure gradient, $G$, between the left inlet boundary and the right outlet boundary. The steady state velocity profile can be analytically computed as:
\begin{equation}
    u_x=\frac{G }{2\mu}y(y-L),
\end{equation}
where $y$ is the vertical coordinate,  $L$ is the side length, and $\mu$ represents the viscosity. The top and bottom boundaries are set to no slip boundary condition. It is noted that besides the steady velocity, the deviatoric
stress $\boldsymbol{T}$ has to be concerned, since it is naturally recovered by the high order moments of the distribution function. The simulation with large relaxation time will affect the accuracy of the recovered Navier-Stokes equation~\cite{reis2020burnett}. In this example, the normal stress, $T_{xx}$, which can be expressed as
\begin{equation}
    T_{xx}=-2\mu\tau(u_x')^2
\end{equation}
decades to zero on the boundary~\cite{reis2020lattice}. 

We first look at the consistency between the numerical results and the analytical solutions shown in Figure~\ref{converge}. In this example, we fixed the Mach number $Ma=0.1/\sqrt{3}$, and the Reynolds number $Re=100$. We evaluate the dimensionless horizontal velocity $u^*_x=u_x/u_{max}$ and normal stress along the $y$ axis when the system achieves the steady state. The velocity profile is highly consistent with the analytical solution. In addition, we do not see obvious spurious oscillations near the boundary region. However, we notice a small difference for stress at the boundary.

We further test the effect of the relaxation time and investigate the convergence of the current boundary as shown in Figure~\ref{slip}. In this test, we fix the maximum velocity by fixing the pressure gradient $G$ but change the relaxation parameter, $\tau$, and
the time interval, $\Delta t$. We first notice that the current model converges to the small time interval, that is, as we decrease the time interval, we will get a smaller "numerical error". However, for the same time interval, as we increase the dimensionless relaxation parameter, a large slip velocity occurs. We learn that since the time marching is solved by SSPRK3 rather than a perfect shift, we can expect a time error at the boundaries. Secondly, the pressure gradient is determined from the viscosity and the relaxation parameter, which may induce a large slip error at the boundary. Besides that, we still notice a more than linear increase in slip velocity which is induced by collision. A further improvement of the boundary condition can be considered by implementing different collision model to obtain a smaller slip velocity.

\subsection{Taylor Green vortex}
\begin{figure}
	\centering
  \includegraphics[width=\linewidth]{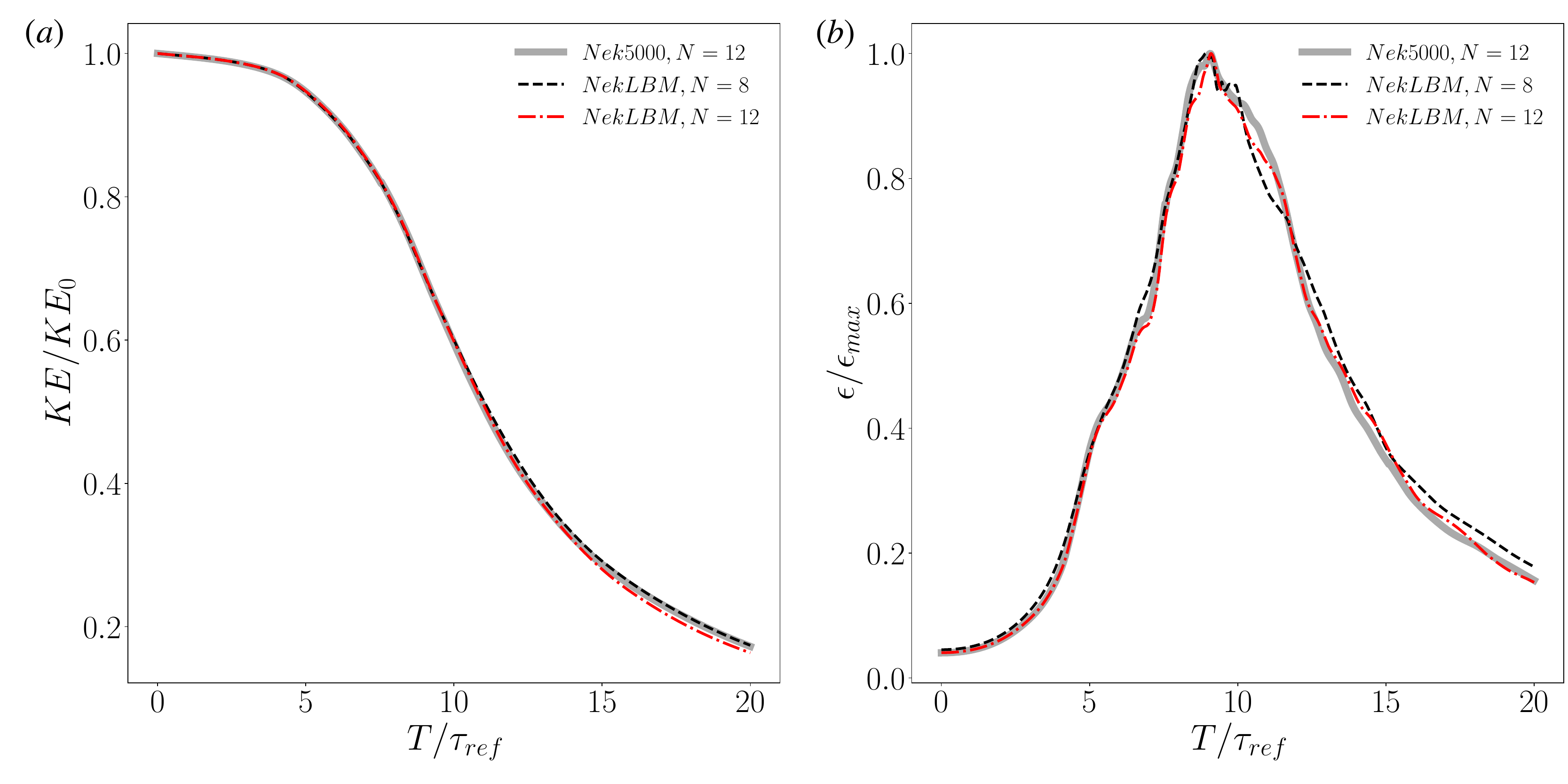}
    \caption{\label{tgv_energy} Evolution of (a) Kinetic energy and (b) energy dissipation during $T/t_0=[0,20]$ of the Taylor Green vortex simulation for the same element number $N_e=16$ on each direction, and different polynomial order $N=8$, $N=12$.
    }
\end{figure}

\begin{figure}
	\centering
  \includegraphics[width=\linewidth]{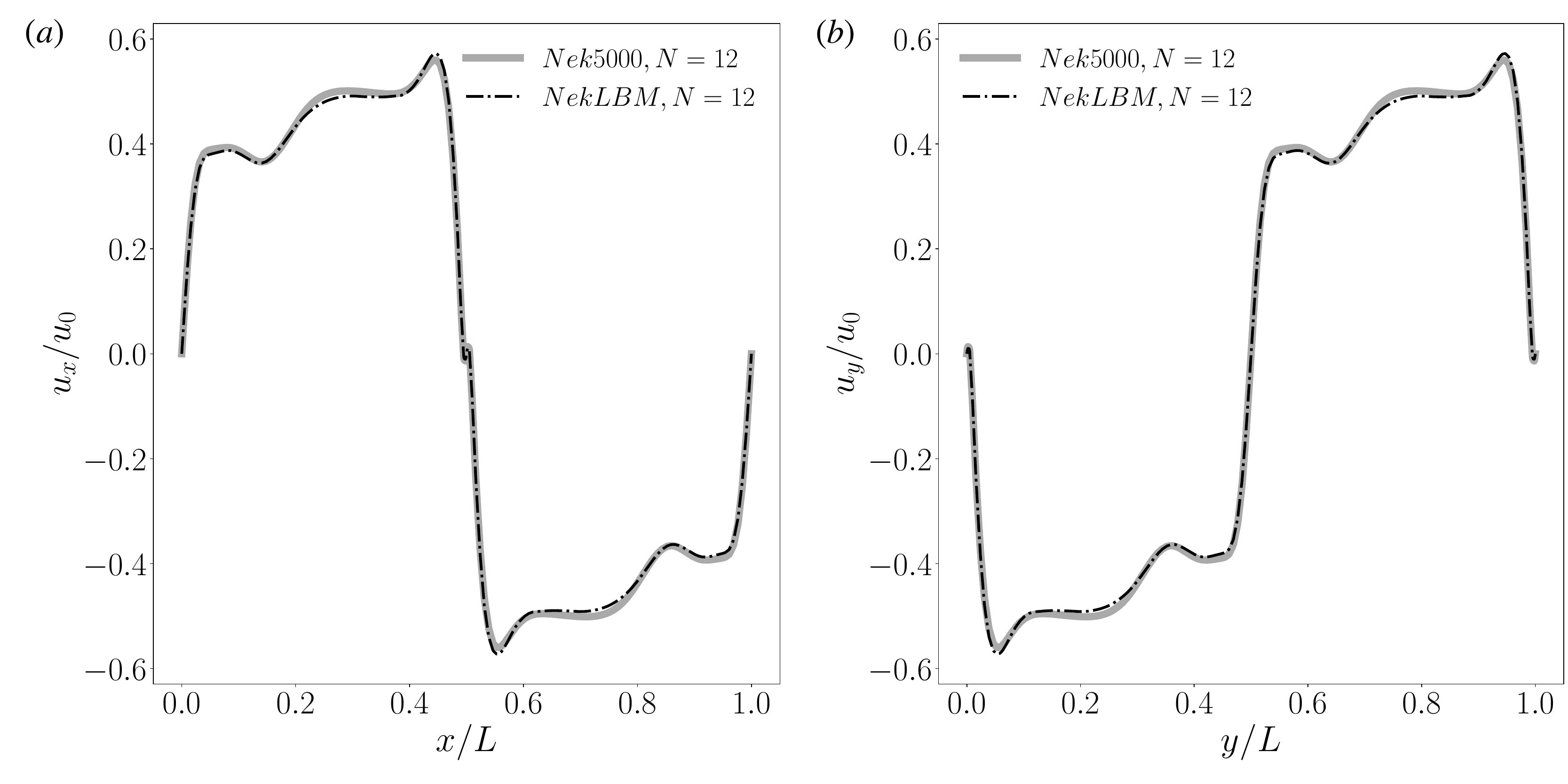}
    \caption{\label{tgv_velo} Velocity in the (a) $x$ ($u_x$) and (b) $y$ ($u_y$) directions along the $x$ and $y$ axial centers at $T/t_0 = 12.11$.
    }
\end{figure}
We then verify the current approach using the 3D Taylor Green vortex case~\cite{brachet1983small} to evaluate the energy dissipation in isotropic turbulence. For this simulation, we use a D3Q19 LBM model~\cite{he1997lattice,kruger2017lattice}. The simulation is initialized with an isotropic velocity profile within a 3D cube with a side length of $L/L_0=2\pi$:
\begin{gather*}\label{TGV_ini}
\begin{split}
&u_x(x,y,z,0)  =u_0 \sin\left(\frac{x}{L_0}\right) \cos\left(\frac{y}{L_0}\right) \cos\left(\frac{z}{L_0}\right), \\
&u_y(x,y,z,0)  =-u_0 \cos\left(\frac{x}{L_0}\right) \sin\left(\frac{y}{L_0}\right) \cos\left(\frac{z}{L_0}\right), \\
&u_z(x,y,z,0)  =0,\\
\end{split}
\end{gather*}
For this simulation, we set $Ma=u_0/c_s=0.1/\sqrt{3}$, $Re=\rho_0 L_0/\nu_0=1600$. The time scale is defined as $t_0=L_0/u_0$. Based on previous research, we evaluate the energy evolution and dissipation rate over the time interval $T/t_0=[0,20]$, comparing the simulation results with Nek5000. Two simulations are performed using the current method, with same element numbers on each direction $N_e=16$, employing the polynomial order $N = 8$ and $N=12$. The first comparison focuses on the kinetic energy $KE=0.5\rho_0\boldsymbol{u}^2$, scaled by the initial kinetic energy $KE_0$. As shown in Figure~\ref{tgv_energy}(a), both simulations exhibit a consistent trend with Nek5000 for $N=12$.

Another comparison, shown in Figure~\ref{tgv_energy}(b), evaluates the dissipation rate as:

\begin{gather*}\label{diss}
\begin{split}
&    \epsilon=2\mu\left[\left(\frac{\partial u_x}{\partial x}\right)^2+\left(\frac{\partial u_y}{\partial y}\right)^2+\left(\frac{\partial u_z}{\partial z}\right)^2-\frac{1}{3}\nabla\cdot\boldsymbol{u}\right]+ \\
&    \mu\left[\left(\frac{\partial u_y}{\partial x}+\frac{\partial u_x}{\partial y}\right)^2+\left(\frac{\partial u_y}{\partial z}+\frac{\partial u_z}{\partial y}\right)^2+\left(\frac{\partial u_x}{\partial z}+\frac{\partial u_z}{\partial x}\right)^2\right]. \\
\end{split}
\end{gather*}
The dissipation rate is scaled by the maximum dissipation, $\epsilon_{max}$. Our approach captures the consistent increase and decrease in the dissipation rate for both cases.

To perform a more sensitive comparison, we analyze the velocity profile at $T/t_0 = 12.11$ along the centerlines in the $x$ and $y$ directions. This comparison, conducted between our approach and Nek5000, is shown in Figure~\ref{tgv_velo}. Our approach with $N=12$ produces highly consistent results compared to Nek5000 with same set up.

\subsection{Pipe Poiseuille flow}
\begin{figure}
	\centering
  \includegraphics[width=\linewidth]{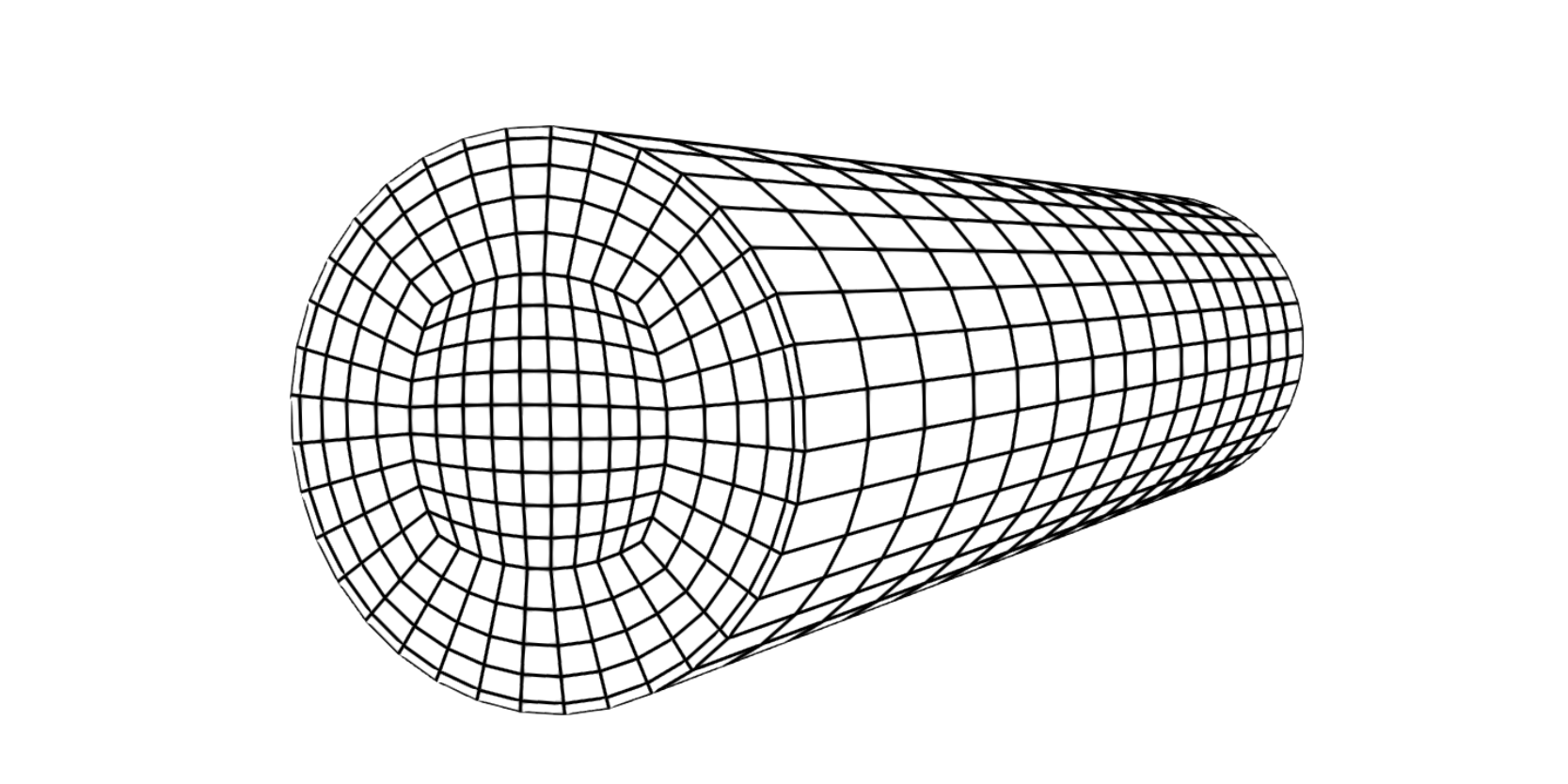}
    \caption{\label{mesh} Elements over the pipe.
    }
\end{figure}
\begin{figure}
	\centering
  \includegraphics[width=\linewidth]{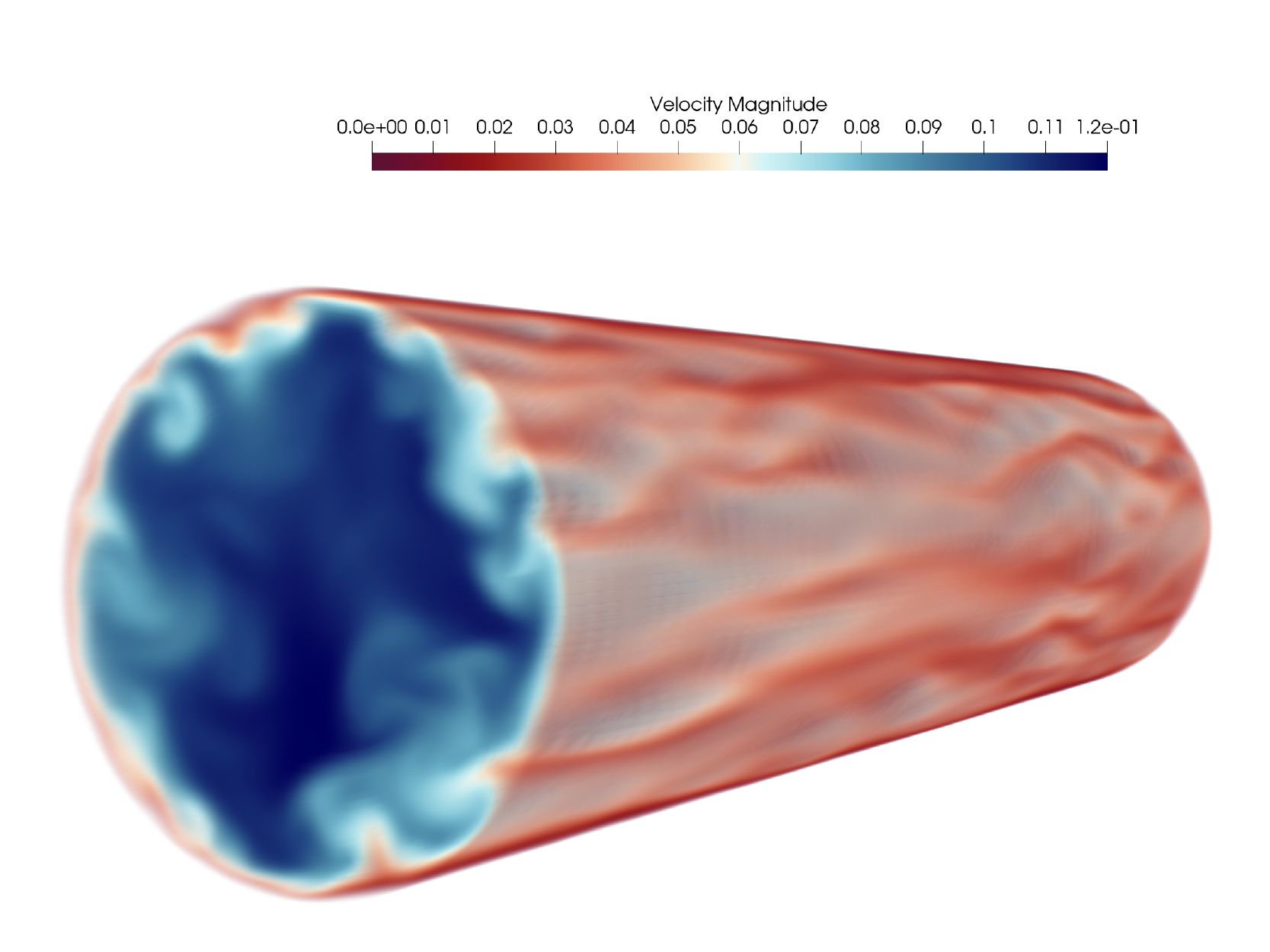}
    \caption{\label{pipe} Volume contours of instantaneous streamwise velocity of a fully-developed turbulent pipe flow at $T/t_0=7.2$ where $Re\approx 5300$. The simulation is conducted with $N_{tot}=3915$, and $N=6$.
    }
\end{figure}
\begin{figure}
	\centering
      \includegraphics[width=\linewidth]{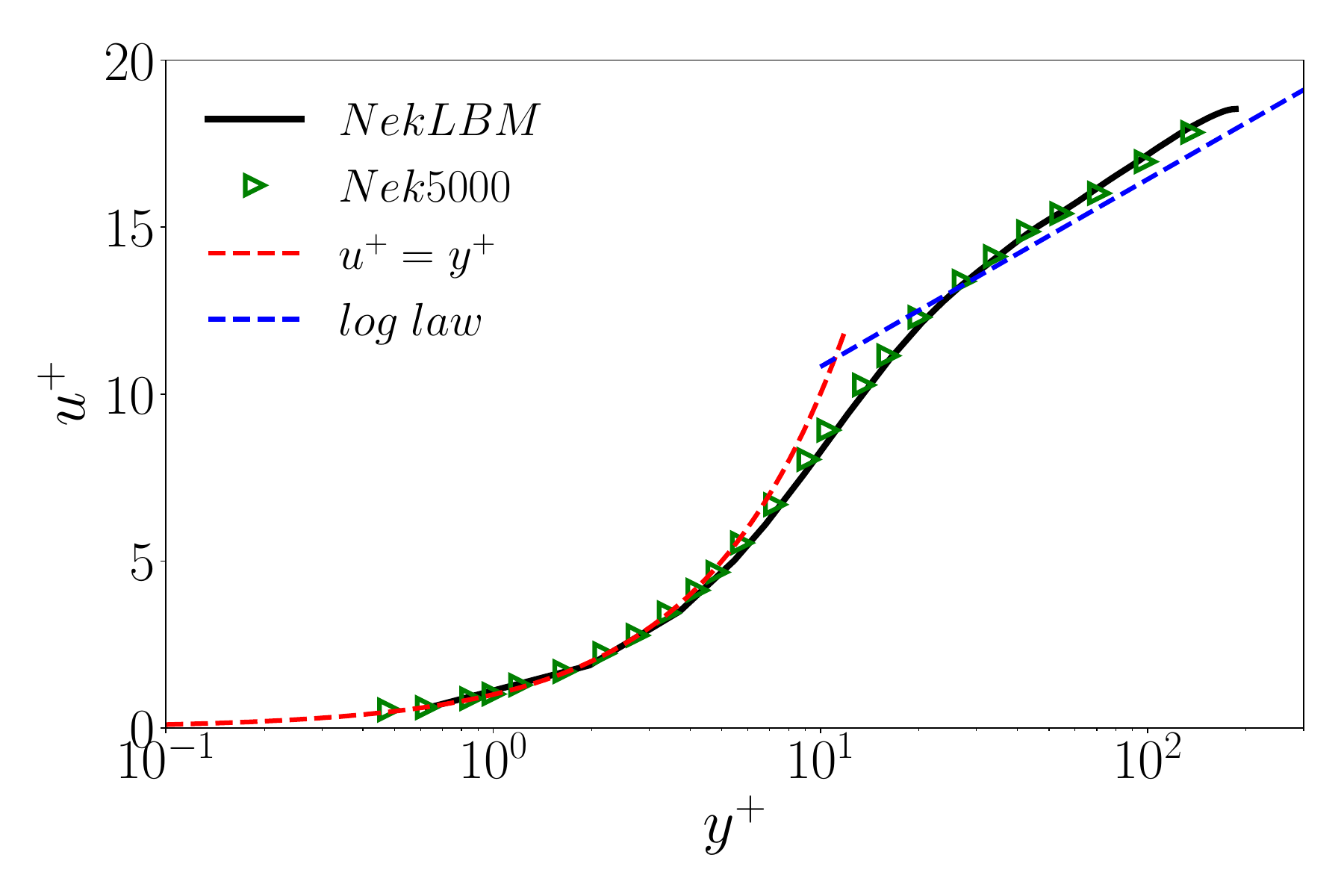}
    \caption{\label{wall} Comparison of the simulation results of turbulent flow in a cylindrical pipe with the law of the wall and Nek5000.
    }
\end{figure}
Finally, we perform a 3D pipe Poiseuille flow simulation using the current scheme. The geometry of the pipe is constructed using GMSH, with the ratio between the diameter and the length $L/D = 4$. An example of the mesh, shown in Figure~\ref{mesh}, consists of $N_e = 3915$ elements. To capture detailed boundary layer information, relatively refined elements are assigned near the edge regions. 

To obtain fully developed turbulence, we initialize the velocity profile which contains the viscous sublayer and inertial sublayer:
\begin{equation}
u^+=\left\{
\begin{aligned}
&y^+, & viscous\ region;\\
&\frac{1}{\kappa}\ln{y^+}+C^+, & inertial\ region.
\end{aligned}
\right.
\end{equation}
Here, the parameter $\kappa=0.41$ represents the Von Kármán constant, and $C^+\approx 5$. In the equations above, $u^+=u_z/u_\tau$ and $u_\tau=\sqrt{\tau_w/\rho}$ are the dimensionless velocity and friction velocity, respectively. The scaled wall coordinate is $y^+=yu_\tau/\nu$, where $\nu$ is the kinematic viscosity. 

We apply the forcing approach suggested in~\cite{peng2018direct} to drive the turbulence. To maintain turbulence, the flow is forced by the body force $(F_x,F_y,F_z)=(0,0,\rho g)$, where $g$ is related to the friction velocity $u_\tau$ as
$g=2u_\tau^2/R$. A further non-uniform, divergence-free force field to the flow during the first three large-eddy turnover times of the simulation, $T/t_0<3$ where $t_0=R/u_\tau$, is given as:

\begin{equation}
    F_r'=-g\kappa B_0\frac{R}{r}\frac{k_z l}{L}\sin{\left(\frac{2\pi t}{P}\right)}\left\{1-\cos{\left[\frac{2\pi(R-r-l_0)}{l}\right]}\right\}
    \cos{\left(k_z\frac{2\pi z}{L}\right)}\cos{(k_\theta\theta)},
\end{equation}
\begin{equation}
    F_\theta'=g(1-\kappa) B_0\frac{k_z}{k_\theta}\frac{2\pi R}{L}\sin{\left(\frac{2\pi t}{P}\right)}\sin{\left[\frac{2\pi(R-r-l_0)}{l}\right]}
    \cos{\left(k_z\frac{2\pi z}{L}\right)}\sin{(k_\theta\theta)},
\end{equation}

\begin{equation}
    F_z'=-g B_0\frac{R}{r}\sin{\left(\frac{2\pi t}{P}\right)}\sin{\left[\frac{2\pi(R-r-l_0)}{l}\right]}
    \sin{\left(k_z\frac{2\pi z}{L}\right)}\cos{(k_\theta\theta)}.
\end{equation}
In the above forcing equations, $k_r$ and $k_\theta$ are wavenumebers of the perturbation in streamwise and azimuthal directions. $P$ is the forcing period. In this case, it is set as $P/t_0=8$. The forcing magnitude is expressed as $B_0=50.0$, and the weighting parameter is denoted as $\kappa=0.5$. The above forcing is added only to the region  $l_0 \leq R-r \leq l_0 +l$ in radial direction, where $l_0=0.2R$, $l=0.4R$. In current scheme, the forcing is then transferred to the cartesian coordinate as,
$(F_x',F_y',F_z')=(F_r'\cos\theta-F_\theta'sin\theta,F_r'sin\theta+F_\theta'\cos{\theta},F_z')$. Finally, the above forcing field is applied to collision and streaming by means of the force split method shown in Eqs.~(\ref{body}) and~(\ref{f_c}). The further validations and details about this forcing approach can be found in~\cite{peng2018direct}.

The simulation is carried out based on the friction Reynolds number $Re_\tau=u_\tau R/\nu=180$, and the fully developed turbulence is obtained based on the Reynolds number $Re=u_0D/\nu\approx5300$, where $u_0$ is the mean velocity.
We present the volume contours of the fully developed turbulence at $T/t_0=7.2$ in Figure~\ref{pipe}. Our goal is to recover the law of the wall, which provides a reliable approximation of the velocity profile in natural streams. We evaluate the wall shear stress, $\tau_w=\mu(\sin{\theta}\partial u_z/\partial y+\cos{\theta}\partial u_z/\partial x)$, using the global average value located at the boundary nodes. The average velocity profile in the axial direction, $u_z$, is approximated by both the time and spatial averages.  For a high Reynolds number, in the inner region of the pipe, the mean velocity parallel to the wall follows a self-similar logarithmic law, and the viscous effects can be neglected. However, near the wall, viscosity significantly affects the flow field.

As shown in Figure~\ref{wall}, the simulation results are consistent with the law of the wall. Closest to the wall, the velocity profile follows the behavior of the near-wall laminar sublayer. This is followed by a transitional buffer region. In the inner region of the pipe far from the wall, we observe the log-law behavior.

\section{Conclusion Remarks}
We present a spectral element continuous Galerkin Lattice Boltzmann method (LBM) to evaluate its capability in solving fluid flow problems in an unstructured mesh. The boundary condition is solved by the flux boundary method which is first applied to discontinuous Galerkin LBM in~\cite{min2011spectral}. Instead of using the perfect shift streaming operator in regular LBM, we solve the streaming step using the SSPRK3 scheme. Unlike previous studies employing the discontinuous Galerkin method, we explore the continuous Galerkin method to further reduce the dissipation introduced by the flux term at element interfaces. To enhance stability, we validate an explicit filter through sensitivity for the double shear layer simulation.

The flux boundary condition is rigorously validated using unsteady Couette flow and planar Poiseuille flow. Both the velocity and stress profiles exhibit excellent agreement with analytical solutions. However, as the relaxation time increases, a noticeable increase in slip velocity is observed. Conversely, reducing the time step results in a relatively smaller slip velocity. Further investigation is required to understand the relationship between slip velocity and relaxation time.

We also validate the dissipation characteristics of the current scheme using the Taylor Green vortex test case. As the number of elements increases, we observe a convergence trend and consistent energy dissipation, which aligns with results from Nek5000.

Finally, we apply the proposed scheme to simulate fully developed turbulent flow in a cylindrical pipe. The simulation is initialized with a constant body forcing between the inlet and outlet. In addition, we apply a non-uniform, divergence-free forcing field to the flow. As the Reynolds number and shear stress reach equilibrium, we compute the average velocity profile and boundary stress. The results recover the turbulent boundary layer profile, accurately capturing the viscous region, buffer region, and log-law region.
\section{Acknowledgment}
This material is based upon work supported by the U.S. Department of Energy (DOE), Office of Nuclear Energy, under Award No. DE-NE0009420, and the National Science Foundation under Grant No. 2344147. This research used resources of the Argonne Leadership Computing Facility, which is a U.S. Department of Energy Office of Science User Facility operated under contract DE-AC02-06CH11357.



 \bibliographystyle{elsarticle-num} 
 \bibliography{cas-refs}





\end{document}